\documentstyle[aps,prd,multicol]{revtex}
\tighten

\def\Lrule{\vspace*{-0.2in}\noindent\vrule width3.4in height.2pt
  depth.2pt \vrule depth0em height.5em}
\def\Rrule{\vspace{-0.1in}\hfill\vrule depth.5em height0pt \vrule
  width3.4in height.2pt depth.2pt\vspace*{-0.1in}}

\begin{document}

\title{Partially embedding of the quantum mechanical
analog of the nonlinear sigma model}

\author{R. Amorim$^a$, J. Barcelos-Neto$^b$ and C. Wotzasek$^c$}

\address{\mbox{}\\
Instituto de F\'{\i}sica\\
Universidade Federal do Rio de Janeiro\\
RJ 21945-970 - Caixa Postal 68528 - Brasil}
\date{\today}

\maketitle
\begin{abstract}
\hfill{\small\bf Abstract\hspace*{1.7em}}\hfill\smallskip
\par
\noindent
We consider the quantum mechanical analog of the nonlinear sigma
model. There are difficulties to completely embed this theory by
directly using the Batalin, Fradkin, Fradkina, and Tyutin (BFFT)
formalism. We show in this paper how the BFFT method can be
conveniently adapted in order to achieve a gauge theory that partially
embeds the model.
\end{abstract}

\smallskip\mbox{}

\begin{multicols}{2}
\section{Introduction}

The interest in embedding of systems with nonlinear constraints has
been started with the works by Banerjee et al. \cite{Banerjee}. The
general and systematic formalism for embedding was developed by
Batalin, Fradkin, Fradkina, and Tyutin (BFFT) \cite{BFFT1,BFFT2} where
systems with second class constraints \cite{Dirac} are transformed
into first class ones, i.e. they are embedded into more general
(gauge) theories. This is achieved with the aid of auxiliary variables
with the general rule such that there is one pair of canonical
variables for each second class constraint to be transformed.

\medskip
The BFFT method is quite elegant and the obtainment of first class
constraints is done in an iterative way. The first correction to the
constraints is linear in the auxiliary variables, the second one is
quadratic, and so on. In the case of systems with just linear
constraints, like chiral-bosons \cite{Floreanini}, one obtains that
just linear corrections are enough to make them first class
\cite{Miao,Barc1}. Here, we mention that the method is equivalent to
express the dynamic quantities by means of shifted coordinates
\cite{Amorim}.

\medskip
However, for systems with nonlinear constraints, the iterative
process may go beyond the first correction. This is a crucial point
for the use of the method. This is so because the first iterative
step may not give a unique solution and one does not know {\it a
priori} what should be the most convenient solution we have to choose
for the second step. There are systems where this choice is very
natural and it is feasible to carry out all the steps. We mention for
example the massive Yang-Mills theory \cite{Barc2}. However, for the
nonlinear sigma-model (and $CP^{N-1}$) not all the solutions of the
first step lead to a solution in the second one \cite{Barc3}. The same
occurs from the second to the third step, and so on, making the
method not feasible to be applied. More than that, in the case of the
nonlinear sigma model one can not assure that these higher order
solutions actually exist \cite{Barc3}. It is important to emphasize
that this is not a problem necessarily related to the method, what may
happen is that there might be no gauge theory that completely embeds
the nonlinear sigma-model.

\medskip
We shall address to this problem in the present paper. We are going to
study the quantum mechanical analog of the nonlinear sigma-model. The
use of the BFFT method in this model also presents similar
difficulties in providing a complete embedding. However, we show how
the method can be conveniently adapted to partially embedding it.

\medskip
Our paper is organized as follows. In Sec. II we make a brief review
of the BFFT method and introduce the general lines of the partially
embedding procedure. In Sec. III, for future comparisons, we discuss
the constraint dynamics of the quantum mechanical analog of the
nonlinear sigma-model, that corresponds to a particle constrained to
move on a $N$-dimensional sphere and show the difficulties we have for
totally embedding it. We develop the partially embedding of this
theory in Sec. IV and discuss the time evolution and the consequences
of the gauge invariance of the model into Sec. V. We left Sec. VI for
some concluding remarks.

\section{Brief review of the BFFT formalism}
\renewcommand{\theequation}{2.\arabic{equation}}
\setcounter{equation}{0}

Let us take a system described by a Hamiltonian $H_c$ in a
phase-space with variables $(q_i, p_i)$ where $i$ runs from 1 to N.
It is also supposed that there exist second class constraints only
since this is the case that will be investigated.  Denoting them by
$T_a$, with $a=1,\dots,M<2N$, we have

\begin{equation}
\bigl\{T_a,\,T_b\bigr\}=\Delta_{ab}\,,
\label{2.1}
\end{equation}

\bigskip\noindent
where $\det(\Delta_{ab})\not=0$.

\medskip
The first objective is to transform these second-class constraints
into first-class ones.  Towards this goal auxiliary variables
$\eta^a$ are introduced, one for each second class constraint (the
connection between the number of constraints and the new variables in
a one-to-one correlation is to keep the same number of the physical
degrees of freedom in the resulting extended theory), which satisfy a
symplectic algebra

\begin{equation}
\label{2.2}
\bigl\{\eta^a,\,\eta^b\bigr\}=\omega^{ab}\,,
\end{equation}

\bigskip\noindent
where $\omega^{ab}$ is a constant quantity with
$\det\,(\omega^{ab})\not=0$.  The first class constraints are now
defined by

\begin{equation}
\label{2.3}
\tilde T_a=\tilde T_a(q,p;\eta)\,,
\end{equation}

\bigskip\noindent
and satisfy the boundary condition

\begin{equation}
\label{2.4}
\tilde T_a(q,p;0)=T_a(q,p)\,.
\end{equation}

\bigskip\noindent
A characteristic of these new constraints is that they are assumed to
be strongly involutive, i.e.

\begin{equation}
\label{2.5}
\bigl\{\tilde T_a,\,\tilde T_b\bigr\}=0\,.
\end{equation}

\bigskip
The solution of~(\ref{2.5}) can be achieved by considering an
expansion of $\tilde T_a$, as

\begin{equation}
\label{2.6}
\tilde T_a=\sum_{n=0}^\infty T_a^{(n)}\,,
\end{equation}

\bigskip\noindent
where $T_a^{(n)}$ is a term of order $n$ in $\eta$.  Compatibility
with the boundary condition~(\ref{2.4}) requires that

\begin{equation}
\label{2.7}
T_a^{(0)}=T_a\,.
\end{equation}

\bigskip\noindent
The replacement of~(\ref{2.6}) into~(\ref{2.5}) leads to a set of
recursive relations, one for each coefficient of $\eta^n$. We
explicitly list the equations for $n =0, 1, 2$,

\end{multicols}
\renewcommand{\theequation}{2.\arabic{equation}}
\Lrule

\begin{eqnarray}
&&\bigl\{T_a^{(0)},T_b^{(0)}\bigr\}_{(q,p)}
+\bigl\{T_a^{(1)},T_b^{(1)}\bigr\}_{(\eta)}=0\,,
\label{2.8a}\\
&&\bigl\{T_a^{(0)},T_b^{(1)}\bigr\}_{(q,p)}
+\bigl\{T_a^{(1)},T_b^{(0)}\bigr\}_{(q,p)}
+\bigl\{T_a^{(1)},T_b^{(2)}\bigr\}_{(\eta)}
+\bigl\{T_a^{(2)},T_b^{(1)}\bigr\}_{(\eta)}=0\,,
\label{2.8b}\\
&&\bigl\{T_a^{(0)},T_b^{(2)}\bigr\}_{(q,p)}
+\bigl\{T_a^{(1)},T_b^{(1)}\bigr\}_{(q,p)}
+\bigl\{T_a^{(2)},T_b^{(0)}\bigr\}_{(q,p)}
+\bigl\{T_a^{(1)},T_b^{(3)}\bigr\}_{(\eta)}
\nonumber\\
&&\phantom{\bigl\{T_a^{(0)},T_b^{(2)}\bigr\}_{(q,p)}}
+\bigl\{T_a^{(2)},T_b^{(2)}\bigr\}_{(\eta)}
+\bigl\{T_a^{(3)},T_b^{(1)}\bigr\}_{(\eta)}=0\,.
\label{2.8c}
\end{eqnarray}

\bigskip
\Rrule
\begin{multicols}{2}
\noindent
The notations $\{,\}_{(q,p)}$ and $\{,\}_{(\eta)}$ represent the parts
of the Poisson bracket $\{,\}$ relative to the variables $(q,p)$ and
$(\eta)$.

\medskip
The above equations are used iteratively to obtain the corrections
$T^{(n)}$ ($n\geq1$).  Equation~(\ref{2.8a}) shall give $T^{(1)}$.
With this result and~(\ref{2.8b}), one calculates $T^{(2)}$, and so
on. Since $T^{(1)}$ is linear in $\eta$ we may write

\begin{equation}
\label{2.9}
T_a^{(1)}=X_{ab}(q,p)\,\eta^b\,.
\end{equation}

\bigskip\noindent
Introducing this expression into~(\ref{2.8a}) and using the
boundary
condition~(\ref{2.4}), as well as~(\ref{2.1})
and~(\ref{2.2}), we get

\begin{equation}
\label{2.10}
\Delta_{ab}+X_{ac}\,\omega^{cd}\,X_{bd}=0\,.
\end{equation}

\bigskip\noindent
We notice that this equation  contains two  unknowns $X_{ab}$ and
$\omega^{ab}$. Usually, first of all $\omega^{ab}$ is chosen in such
a way that the new variables are unconstrained. It is opportune to
mention that it is not always possible to make such a choice
\cite{Barc1}. In consequence, the consistency of the method requires
the introduction of other new variables in order to transform these
constraints also into first-class. This may lead to an endless
process. However, it is important to emphasize that $\omega^{ab}$ can
be fixed anyway.

\medskip
After fixing $\omega^{ab}$, we pass to consider the coefficients
$X_{ab}$. They cannot be obtained  unambiguously since, even after
fixing $\omega^{ab}$, expression (\ref{2.10}) leads to less equations
than variables. The choice of $X$'s has therefore to be done in a
convenient way \cite{Banerjee}.

\medskip
The knowledge of $X_{ab}$ permits us to obtain $T_a^{(1)}$. If
$X_{ab}$ does not depend on $(q,p)$, it is easily seen that
$T_a+T_a^{(1)}$ is already strongly involutive and we succeed in
obtaining $\tilde T_a$.  This is what happens for systems with linear
constraints. For nonlinear constraints, on the other hand, $X_{ab}$
becomes variable dependent which necessitates the analysis to be
pursued beyond the first iterative step.  All the subsequent
corrections must be explicitly computed, the knowledge of $T_a^{(n)}
(n=0, 1, 2,...n) $ leading to the evaluation of $T_a^{(n+1)}$ from
the recursive relations.  Once again the importance of choosing the
proper  solution for $X_{ab}$ becomes apparent otherwise the series
of corrections cannot be put in a closed form and the expression for
the involutive constraints becomes unintelligible and uninteresting.

\medskip
Another point in the Hamiltonian formalism is that any dynamic
function $A(q,p)$ (for instance, the Hamiltonian) has also to be
properly modified in order to be strongly involutive with the
first-class constraints $\tilde T_a$. Denoting the modified quantity
by $\tilde A(q,p;\eta)$, we then have

\begin{equation}
\label{2.11}
\bigl\{\tilde T_a,\,\tilde A\bigr\}=0\,.
\end{equation}

\bigskip\noindent
In addition, $\tilde A$ has also to satisfy  the boundary
condition

\begin{equation}
\label{2.12}
\tilde A(q,p;0)=A(q,p)\,.
\end{equation}

\bigskip
To obtain $\tilde A$ an expansion analogous to (\ref{2.6}) is
considered,

\begin{equation}
\label{2.13}
\tilde A=\sum_{n=0}^\infty A^{(n)}\,,
\end{equation}

\bigskip\noindent
where $A^{(n)}$ is also a term of order $n$ in $\eta$'s.
Consequently, compatibility with~(\ref{2.12}) requires that

\begin{equation}
\label{2.14}
A^{(0)}=A\,.
\end{equation}

\bigskip\noindent
The combination of~(\ref{2.6}),~(\ref{2.11}) and~(\ref{2.13}) gives

\end{multicols}
\Lrule

\begin{eqnarray}
&&\bigl\{T_a^{(0)},A^{(0)}\bigr\}_{(q,p)}
+\bigl\{T_a^{(1)},A^{(1)}\bigr\}_{(\eta)}=0\,,
\label{2.15a}\\
&&\bigl\{T_a^{(0)},A^{(1)}\bigr\}_{(q,p)}
+\bigl\{T_a^{(1)},A^{(0)}\bigr\}_{(q,p)}
+\bigl\{T_a^{(1)},A^{(2)}\bigr\}_{(\eta)}
\nonumber\\
&&\phantom{\bigl\{T_a^{(0)},A^{(0)}\bigr\}_{(q,p)}}
+\bigl\{T_a^{(2)},A^{(1)}\bigr\}_{(\eta)}=0\,,
\label{2.15b}\\
&&\bigl\{T_a^{(0)},A^{(2)}\bigr\}_{(q,p)}
+\bigl\{T_a^{(1)},A^{(1)}\bigr\}_{(q,p)}
+\bigl\{T_a^{(2)},A^{(0)}\bigr\}_{(q,p)}
\nonumber\\
&&\phantom{\bigl\{T_a^{(0)},A^{(0)}\bigr\}_{(q,p)}}
+\bigl\{T_a^{(1)},A^{(3)}\bigr\}_{(\eta)}
+\bigl\{T_a^{(2)},A^{(2)}\bigr\}_{(\eta)}
\nonumber\\
&&\phantom{\bigl\{T_a^{(0)},A^{(0)}\bigr\}_{(q,p)}}
+\bigl\{T_a^{(3)},A^{(1)}\bigr\}_{(\eta)}=0\,,
\label{2.15c}
\end{eqnarray}

\bigskip

\noindent
which correspond to the coefficients of the powers $\eta^0$,
$\eta^1$, $\eta^2$, etc., respectively. The expression~(\ref{2.15a})
above gives us $A^{(1)}$

\begin{equation}
\label{2.16}
A^{(1)}=-\,\eta^a\,\omega_{ab}\,X^{bc}\,
\bigl\{T_c,\,A\bigr\}\,,
\end{equation}

\bigskip\noindent
where $\omega_{ab}$ and $X^{ab}$ are the inverses of $\omega^{ab}$
and $X_{ab}$.

\medskip
It was earlier seen that $T_a+T^{(1)}_a$ was strongly involutive if
the coefficients $X_{ab}$ do not depend on $(q,p)$. However, the same
argument does not necessarily apply in this case. Usually we have to
calculate other corrections to obtain the final $\tilde A$. Let us
discuss how this can be systematically done. The correction $A^{(2)}$
comes from equation~(\ref{2.15b}), that we conveniently rewrite as

\begin{equation}
\label{2.17}
\bigl\{T_a^{(1)},\,A^{(2)}\bigr\}_{(\eta)}
=-\,G_a^{(1)}\,,
\end{equation}

\bigskip

\noindent
where
\begin{equation}
\label{2.18}
G_a^{(1)}=\bigl\{T_a,\,A^{(1)}\bigr\}_{(q,p)}
+\bigl\{T_a^{(1)},\,A\bigr\}_{(q,p)}
+\bigl\{T_a^{(2)},\,A^{(1)}\bigr\}_{(\eta)}\,.
\end{equation}
\bigskip

\noindent
Thus

\begin{equation}
\label{2.19}
A^{(2)}=-{1\over2}\,\eta^a\,\omega_{ab}\,X^{bc}\,G_c^{(1)}\,.
\end{equation}

\bigskip\noindent
In the same way, other terms can be obtained. The final general
expression reads

\begin{equation}
\label{2.20}
A^{(n+1)}=-{1\over n+1}\,\eta^a\,\omega_{ab}\,X^{bc}\,G_c^{(n)}\,,
\end{equation}

\bigskip


\noindent
where

\begin{equation}
\label{2.21}
G_a^{(n)}=\sum_{m=0}^n\bigl\{T_a^{(n-m)},\,A^{(m)}\bigr\}_{(q,p)}
+\sum_{m=0}^{n-2}\bigl\{T_a^{(n-m)},\,A^{(m+2)}\bigr\}_{(\eta)}
+\bigl\{T_a^{(n+1)},\,A^{(1)}\bigr\}_{(\eta)}\,.
\end{equation}

\bigskip

\begin{multicols}{2}
The partially embedding procedure we are going to apply consists in
transforming into first-class just part of the constraints. Let us
suppose we take $M^\prime$  among the $M$ second-class constraints to
be converted into first-class. There are two general steps to be done.
The first one is to achieve the strong involutive algebra for these
constraints, namely,

\begin{equation}
\{\tilde T_{a'},\tilde T_{b'}\}=0
\hspace{1cm} a',b'=1,\dots,M^\prime
\label{2.22}
\end{equation}

\bigskip\noindent
by introducing $M^\prime$ variables $\eta^{a'}$ of auxiliary variables
and the same steps of the BFFT formalism. Consequently, in the
partially embedding formalism we work with less auxiliary variables
than in the full procedure. Of course, the choice of what constraints
we intend to convert into first-class may be a crucial point for the
success of the method.

\medskip
The fact of having achieved a strong involutive algebra for some of
the constraints does not necessarily means that these constraints are
first-class. We have also to convert the remaining $M-M^\prime$ ones
in order to have involutive algebras with all the $\tilde
T_{a'}$. This is the second general step we have talked above and it
is achieved by treating the remaining constraints in a similar way of
the quantity $A$ that appears in (\ref{2.11}). Denoting the remaining
second-class constraints by $T_{a"}$, with $a"=1,\dots M-M^\prime$, we
would have to obtain $\tilde T_{a"}(q,p,\eta^\prime)$ in the same
lines that $\tilde A$ was done in the BFFT formalism. In this way

\begin{equation}
\{\tilde T_{a'},\tilde T_{a"}\}=0\,,
\label{2.23}
\end{equation}

\bigskip\noindent
but, in general, the matrix $\Delta_{a"b"}=\{\tilde T_{a"},\tilde
T_{b"}\}$ will be nonsingular, which means that the $\tilde T_{a"}$'s
remain second-class

\medskip
Finally we mention that the strong involutive Hamiltonian are obtained
in the same way as $\tilde T_{a"}$. Further details will be displayed
when we use the formalism in the example we are going to consider.

\section{The model and difficulties for totally embedding}
\renewcommand{\theequation}{3.\arabic{equation}}
\setcounter{equation}{0}

\bigskip
The motion of a particle on a $N$-dimensional sphere of radius 1 is
described by the Lagrangian

\begin{equation}
L=\frac{1}{2}\,\dot q_i\dot q_i
+\frac{1}{2}\,\lambda\,(q_iq_i-1)\,.
\label{3.1}
\end{equation}

\bigskip
\noindent
As can be easily verified, the Euler-lagrange equations for $\lambda$
and $q_i$ are respectively given by

\begin{eqnarray}
&&q^2-1=0\,,
\nonumber\\
&&\ddot q_i-\lambda q_i=0\,.
\label{3.2}
\end{eqnarray}

\bigskip\noindent
In the first equation above we are using a short notation. We do it
from now on where there is no misunderstanding. Derivating the first
of the equations above twice with respect to the evolution parameter
and using the second equation, we see that $\lambda=-\,\dot q^2/q^2$,
and so we get the expected equation of motion for a particle moving on
a $N$-dimensional sphere:

\begin{equation}
\ddot q_i=-\,\frac{\dot q^2}{q^2}\,q_i\,.
\label{3.3}
\end{equation}

\bigskip
Let us now consider this theory in the canonical formalism. The
canonical Hamiltonian corresponding to (\ref{3.1}) reads

\begin{equation}
H_c=\frac{1}{2}\,p^2-\frac{1}{2}\,\lambda\,(q^2-1)\,,
\label{3.4}
\end{equation}

\bigskip\noindent
where $p_i$ is the canonical conjugated momentum to $q_i$. Using the
Dirac constraint formalism \cite{Dirac} we obtain that the constraints
of this theory are

\begin{eqnarray}
T_1&=&q^2-1\,,
\nonumber\\
T_2&=&q.p\,,
\nonumber\\
T_3&=&p_\lambda\,,
\nonumber\\
T_4&=&\lambda\,q^2+p^2\,,
\label{3.5}
\end{eqnarray}

\bigskip\noindent
where $p_\lambda$ is the canonical conjugated momentum to $\lambda$.
These constraints are second class. In fact, for the antisymmetric
quantities $\Delta_{ab}$ given by (\ref{2.1}), we have

\begin{eqnarray}
\Delta_{12}&=&2\,q^2\,,
\nonumber\\
\Delta_{14}&=&4\,q.p\,,
\nonumber\\
\Delta_{24}&=&-2\,\lambda\,q^2+2\,p^2\,,
\nonumber\\
\Delta_{34}&=&-\,q^2\,,
\label{3.6}
\end{eqnarray}

\bigskip\noindent
and one can verify that the matrix $\Delta=(\Delta_{ab})$ is regular.

\medskip
The Dirac brackets between any two quantities $A$ and $B$ is
constructed in the usual way,

\begin{equation}
\{A,B\}_D=\{A,B\}-\{A,T_a\}\Delta^{-1}_{ab}\{T_b,B\}
\label{3.7}
\end{equation}

\bigskip\noindent
and since any quantity has null Dirac brackets with any one of the
$T$'s, the time evolution  generated by $H_c$, or by any of its
extensions by adding to it terms such as $\lambda_a T_a$, gives the
same result. Actually, by considering the form of the constraints, we
note that (\ref{3.4}) can be written as

\begin{equation}
H_c=-\,\lambda\,T_1+\frac{1}{2}\,T_4-\frac{1}{2}\,\lambda\,,
\label{3.8}
\end{equation}

\bigskip\noindent
and so, under Dirac brackets, the dynamics of the system is generated
just by $-\frac{1}{2}\,\lambda$. As can be verified,

\begin{eqnarray}
&&\dot q_i=-\,\frac{1}{2}\,\{q_i,\lambda\}_D
=\frac{1}{q^2}\,p_i\,,
\nonumber\\
&&\dot p_i=-\,\frac{1}{2}\,\{p_i,\lambda\}_D
=-\frac{\lambda}{q^2}\,q_i\,,
\nonumber\\
&&\dot\lambda=-\,\frac{1}{2}\,\{\lambda,\lambda\}_D=0\,,
\label{3.9}
\end{eqnarray}

\bigskip\noindent
which give, on the constraint surface, the same dynamics as the one
generated by the Euler-Lagrange equations (\ref{3.2}).

\medskip
Let us now try to use the full BFFT method for the present theory in
order to see the difficulties for embedding it. The use of the method
requires the introduction of four coordinates $\eta^a$. We consider
them such that $\eta^3$ and $\eta^4$ are the momenta conjugated to
$\eta^1$ and $\eta^2$, respectively. Hence, the matrix $(\omega^{ab})$
given by (\ref{2.2}) reads

\begin{equation}
(\omega^{ab})=\left(\begin{array}{cccc}
0&0&1&0\\
0&0&0&1\\
-1&0&0&0\\
0&-1&0&0\\
\end{array}\right)\,.
\label{3.10}
\end{equation}

\bigskip\noindent
The combination of (\ref{2.11}), (\ref{2.12}), (\ref{3.3}), and
(\ref{3.10}) leads to the set of equations

\begin{eqnarray}
&&X_{11}X_{23}+X_{12}X_{24}-X_{13}X_{21}-X_{14}X_{22}=-2q^2\,,
\nonumber\\
&&X_{11}X_{33}+X_{12}X_{34}-X_{13}X_{31}-X_{14}X_{32}=0\,,
\nonumber\\
&&X_{11}X_{43}+X_{12}X_{44}-X_{13}X_{41}-X_{14}X_{42}=-4q.p\,,
\nonumber\\
&&X_{21}X_{33}+X_{22}X_{34}-X_{23}X_{31}-X_{24}X_{32}=0\,,
\nonumber\\
&&X_{21}X_{43}+X_{22}X_{44}-X_{23}X_{41}-X_{24}X_{42}=
2\lambda q^2-2p^2\,,
\nonumber\\
&&X_{31}X_{43}+X_{32}X_{44}-X_{33}X_{41}-X_{34}X_{42}=q^2\,.
\label{3.11}
\end{eqnarray}

\bigskip\noindent
The system above cannot be univocally solved. It contains sixteen
variables in just six equations. In cases like this we examine the
possibility of figuring out a solution where the first linear
correction for the constrains could lead to a strongly involutive
algebra. This is achieved if besides Eq. (\ref{2.12}) the equations

\begin{eqnarray}
&&\{T_a,X_{bc}\}+\{X_{ac},T_b\}=0
\nonumber\\
&&\{X_{ac},X_{bd}\}+\{X_{ad},X_{bc}\}=0
\label{3.12}
\end{eqnarray}

\bigskip\noindent
are also satisfied. Since the coefficients $X_{ab}$ depend on
coordinates and momenta, there is no choice where this can be
achieved.

\medskip
In order to see the problem of going to the next steps of the method,
let us make a choice that solves (\ref{3.11}):

\begin{equation}
\begin{array}{llll}
X_{11}=0,&X_{12}=0,&X_{13}=2,&X_{14}=q.p,\\
X_{21}=q^2,&X_{22}=0,&X_{23}=0,&X_{24}=\frac{1}{2}p^2,\\
X_{31}=0,&X_{32}=0,&X_{33}=0,&X_{34}=-\frac{1}{4}q^2,\\
X_{41}=0,&X_{42}=4,&X_{43}=2\lambda,&X_{44}=0.\\
\end{array}
\label{3.13}
\end{equation}

\bigskip\noindent
With this choice we have the following first-order correction for the
constraints

\begin{eqnarray}
T_1^{(1)}&=&2\,\eta^3+q.p\,\eta^4\,,
\nonumber\\
T_2^{(1)}&=&q.q\,\eta^1+\frac{1}{2}\,p.p\,\eta^4\,,
\nonumber\\
T_3^{(1)}&=&-\frac{1}{4}\,q.q\,\eta^4\,,
\nonumber\\
T_4^{()1}&=&4\,\eta^2+2\,\lambda\,\eta^3\,.
\label{3.14}
\end{eqnarray}

\bigskip
We now have to consider these quantities into expression
(\ref{2.9}),which leads to the following set of equations

\end{multicols}
\renewcommand{\theequation}{3.\arabic{equation}}
\Lrule

\begin{eqnarray}
&&4\frac{\partial T_2^{(2)}}{\partial\eta^1}
+2q.p\frac{\partial T_2^{(2)}}{\partial\eta^2}
+q.q\frac{\partial T_1^{(2)}}{\partial\eta^3}
-p.p\frac{\partial T_1^{(2)}}{\partial\eta^2}
=4q.p\eta^4\,,
\nonumber\\
&&8\frac{\partial T_3^{(2)}}{\partial\eta^1}
+4q.p\frac{\partial T_3^{(2)}}{\partial\eta^2}
+q.q\frac{\partial T_1^{(2)}}{\partial\eta^2}
=0\,,
\nonumber\\
&&2\frac{\partial T_4^{(2)}}{\partial\eta^1}
+q.p\frac{\partial T_4^{(2)}}{\partial\eta^2}
+4\frac{\partial T_1^{(2)}}{\partial\eta^4}
-2\lambda\frac{\partial T_1^{(2)}}{\partial\eta^1}
=2\eta^4(p.p-\lambda q.q)\,,
\nonumber\\
&&4q.q\frac{\partial T_3^{(2)}}{\partial\eta^3}
-2p.p\frac{\partial T_3^{(2)}}{\partial\eta^2}
-q.q\frac{\partial T_2^{(2)}}{\partial\eta^4}
=-2q.q\eta^4\,,
\nonumber\\
&&2q.q\frac{\partial T_4^{(2)}}{\partial\eta^3}
-p.p\frac{\partial T_4^{(2)}}{\partial\eta^2}
-8\frac{\partial T_2^{(2)}}{\partial\eta^4}
+4\lambda\frac{\partial T_2^{(2)}}{\partial\eta^1}
=4q.p(\lambda\eta^4-2\eta^1)\,,
\nonumber\\
&&q.q\frac{\partial T_4^{(2)}}{\partial\eta^2}
-16\frac{\partial T_3^{(2)}}{\partial\eta^4}
+8\lambda\frac{\partial T_3^{(2)}}{\partial\eta^1}
=4(\eta^3+q.p\eta^4)\,.
\label{3.15}
\end{eqnarray}

\bigskip

\begin{multicols}{2}
\noindent
As one observes, this system may have many solutions. This can be
verified if one writes $T_a^{(2)}$ as $X_{abc}\,\eta^b\eta^c$. So each
$T_a^{(2)}$ has sixteen terms and the six equations (\ref{3.15}) will
involve ninety six quantities to be fixed. For any choice we make,
this problem will be enlarged and enlarged at each step of the method.

\medskip
We then observe that it is not feasible to infer what is the general
rule for the corrections and, consequently, this discard any
possibility of obtaining a closed solution. We may conclude that the
use of full BFFT method to this problem is very tedious and
uninteresting.

\vspace{1cm}
\section{Using the partially embedding formalism}
\renewcommand{\theequation}{4.\arabic{equation}}
\setcounter{equation}{0}

\bigskip
Considering again the set of constraints (\ref{3.5}), let us just
convert $T_1$ and $T_2$ into first class and let $T_3$ and $T_4$ as
second class constraints. Then instead of two pair of canonical
coordinated we introduce just one, that we simply denote by $\eta^1=
\eta$ and $\eta^2=\pi$. From the solutions of the first step of the
BFFT method we make an analogous choice of Banerjee et al.
\cite{Banerjee}

\begin{eqnarray}
&&\tilde T_1=T_1+T_1^{(1)}=q^2-1+2\eta\,,
\nonumber\\
&&\tilde T_2=T_2+T_2^{(1)}=q.p-\pi q^2\,,
\label{4.1}
\end{eqnarray}

\bigskip\noindent
giving

\begin{equation}
\{\tilde T_1,\tilde T_2\}=0\,.
\label{4.3}
\end{equation}

\bigskip
It is opportune to mention that this choice, which would correspond to
$X_{11}=2$ and $X_{23}=-q.q$ of the full BFFT method is not compatible
with any solution for the set of equations given by (\ref{3.11}).

\medskip
Of course, the result given by (\ref{4.3}) does not necessarily means
that $\tilde T_1$ and $\tilde T_2$ are first class. They also have to
have zero Poisson brackets, on the constraint surface, with the
remaining constraints. We notice that this is actually true for the
constraint $T_3$, but it is not for $T_4$. Let us then conveniently
modify the constraint $T_4$ in order to have zero Poisson brackets
with $\tilde T_1$ and $\tilde T_2$. As it was mentioned in Sec. II,
this can be achieved in the same framework of the BFFT formalism by
taking $T_4$ as the quantity $A$ of Eq. (\ref{2.11}). So, the general
expression for the first correction for $T_4$ should be

\begin{equation}
T_4^{(1)}=-\,\eta^{a'}\omega_{a'b'}X^{bc}\,\{T_{c'},T_4\}\,,
\label{4.4}
\end{equation}

\bigskip\noindent
where the indices $a^\prime,b',c'=1,2$ just correspond to the
constraints $\tilde T_1$ and $\tilde T_2$, and $\omega_{a'b'}$ and
$X^{a'b'}$ are the inverse of $\omega^{a'b'}$ and $X_{a'b'}$
respectively. Considering expressions (\ref{3.4}) and (\ref{3.5}) we
have

\begin{equation}
\omega_{a'b'}=\left(\begin{array}{cc}
0&-1\\
1&0\\
\end{array}\right)\,,
\hspace{1cm}
X^{a'b'}=\left(\begin{array}{cc}
\frac{1}{2}&0\\
0&-\frac{1}{q^2}
\end{array}\right)\,.
\label{4.5}
\end{equation}

\bigskip\noindent
Hence, the first correction for $T_4$ reads

\begin{equation}
T_4^{(1)}=2\Bigl(\lambda-\,\frac{p^2}{q^2}\Bigr)\eta
-2p.q\,\pi\,.
\label{4.6}
\end{equation}

\bigskip\noindent
Using (\ref{2.17})--(\ref{2.21}) we calculate other corrections for
$T_4$. We list some of them below

\begin{eqnarray}
T_4^{(2)}&=&\Bigl(\frac{2p_i}{q^2}\,\eta+q_i\,\pi\Bigr)^2\,,
\nonumber\\
T_4^{(3)}&=&-\,\frac{2\eta}{q^2}\,
\Bigl(\frac{2p_i}{q^2}\,\eta+q_i\,\pi\Bigr)^2\,,
\nonumber\\
T_4^{(4)}&=&\frac{4\eta^2}{q^4}\,
\Bigl(\frac{2p_i}{q^2}\,\eta+q_i\,\pi\Bigr)^2\,.
\label{4.7}
\end{eqnarray}

\bigskip\noindent
From these results we can infer that the general correction
$T_4^{(n)}$, for $n\geq2$, should be

\begin{equation}
T_4^{(n)}=\Bigl(-\,\frac{2\eta}{q^2}\Bigr)^{n-2}\,
\Bigl(\frac{2p_i}{q^2}\,\eta+q_i\,\pi\Bigr)^2\,.
\label{4.8}
\end{equation}

\bigskip
We then see that the partially embedding procedure, contrarily to the
use of the full BFFT method, permitted us to infer the general rule
for all the corrections. More than that, we can also show that the sum
of all these terms to obtain $\tilde T_4$ can be cast in a closed
form,

\begin{eqnarray}
\tilde T_4&=&\lambda\,q^2+p^2+2\lambda\,\eta
-2\,p_i\Bigl(\frac{p_i}{q^2}\,\eta+q_i\,\pi\Bigr)
\nonumber\\
&&+\Bigl(\frac{2p_i}{q^2}\,\eta+q_i\,\pi\Bigr)^2\,
\sum_{n=0}^\infty\Bigl(-\,\frac{2\eta}{q^2}\Bigr)^n\,,
\nonumber\\
&=&\lambda\,q^2+p^2
+2\,\Bigl(\lambda-\frac{p^2}{q^2}\Bigr)\,\eta
-2\,p.q\,\pi
\nonumber\\
&&+\Bigl(\frac{2p_i}{q^2}\,\eta+q_i\,\pi\Bigr)^2\,
\Bigl(1+\frac{2\eta}{q^2}\Bigr)^{-1}\,.
\label{4.9}
\end{eqnarray}

\bigskip\noindent
This constraint can be further rewritten as

\begin{equation}
\tilde T_4=\Bigl(1+\frac{2\eta}{q^2}\Bigr)^{-1}
\bigl(p_i-\pi q_i\bigr)^2
+\lambda\bigl(q^2+2\eta\bigr)\,.
\label{4.10}
\end{equation}

\bigskip
It is just a matter of algebraic work to check that $\tilde T_1$ and
$\tilde T_2$ are actually first class, whereas $\tilde
T_3=T_3=p_\lambda$ and $\tilde T_4$ are second class.

\medskip
Using the partially embedding procedure in terms of the first-class
constraints $\tilde T_1$ and $\tilde T_2$ we directly obtain the
partially embedding Hamiltonian $\tilde H_c$ that resembles the form
of the Hamiltonian $H_c$ given by (\ref{3.8}), namely

\begin{equation}
\tilde H_c=-\,\lambda\,\tilde T_1
+\frac{1}{2}\,\tilde T_4-\frac{1}{2}\,\lambda\,.
\label{4.11}
\end{equation}

\bigskip\noindent
It is important to emphasize that it generates a consistent time
evolution for each one of the constraints $\tilde T_a$.

\section{Time evolution and gauge invariance}
\renewcommand{\theequation}{5.\arabic{equation}}
\setcounter{equation}{0}

The first order Lagrangian for the theory described in the last
section reads

\begin{equation}
\tilde L=p\cdot\dot q+\pi\dot\eta+p_\lambda\dot\lambda-\tilde H\,,
\label{5.2}
\end{equation}

\bigskip\noindent
where $\tilde H$ is the total Hamiltonian

\begin{equation}
\tilde H=\tilde H_c+\lambda_{a'}\tilde T_{a'}\,.
\label{5.3}
\end{equation}

\bigskip
This theory must be invariant under the transformations generated by
the first-class constraints. Considering that $y$ represents any one
of the canonical coordinates of the system, we have

\begin{equation}
\delta y=\epsilon_{a'}\{y,\tilde T_{a'}\}_D\,,
\label{5.4}
\end{equation}

\bigskip\noindent
where $\epsilon_{a'}$ is the parameter characteristic of the gauge
transformation generated by the first-class constraint
$\tilde T_{a'}$. The presence of the Dirac brackets here is to
consistently eliminate the constraints $T_{a"}$. We thus obtain

\begin{eqnarray}
\delta q_i&=&\epsilon_2\,q_i\,,
\nonumber\\
\delta p_i&=&-\,2\,\epsilon_1q_i-\epsilon_2(p_i-2\pi q_i)\,,
\nonumber\\
\delta\eta&=&-\,\epsilon_2q^2\,,
\nonumber\\
\delta\pi&=&-\,2\,\epsilon_1\,,
\nonumber\\
\delta\lambda&=&0\,,
\nonumber\\
\delta p_\lambda&=&0\,.
\label{5.5}
\end{eqnarray}

\bigskip\noindent
The gauge invariance of the corresponding action is then achieved if
the Lagrange multipliers $\lambda_{a'}$ transform as

\begin{equation}
\delta\lambda_{a'}=-\,\dot\epsilon_{a'}\,.
\label{5.6}
\end{equation}

\bigskip
Let us finally consider the equations of motion generated by the total
Hamiltonian (\ref{5.3}). An important point regarding the embedding
procedure is that the obtained theory, even though having more
symmetries than the initial one, does not change its physics. In other
words, the theory described by the total Hamiltonian $\tilde H$ must
describe a particle moving on a sphere of radius one. If the partially
embedding we have developed till now makes sense, this point has
necessarily to be verified.

\medskip
The general expression for the time evolution of any canonical
quantity, on the constraint surface, is

\begin{eqnarray}
\dot y&=&\{y,\tilde H\}_D\,,
\nonumber\\
&=&-\,\frac{1}{2}\{y,\lambda\}_D
+(\lambda-\lambda_1)\{\tilde T_1,y\}_D
-\lambda_2\{\tilde T_2,y\}_D\,.
\label{5.7}
\end{eqnarray}

\bigskip\noindent
This leads, after some simplifications, to the equations of motion

\end{multicols}
\renewcommand{\theequation}{5.\arabic{equation}}

\Lrule

\begin{eqnarray}
\dot q_i&=&\Bigl(2\tilde T_1+1\Bigr)
\frac{q^2}{(q^2+2\eta)^2}\Bigl(p_i-2\pi q_i\Bigr)
-\lambda^2q_i\,,
\label{5.8}\\
\dot p_i&=&\pi\dot q_i+\dot\pi q_i
+\lambda^2(p_i-\pi q_i)
-(2\tilde T_1+1)\Bigl(\frac{p_k+\pi q_k}{q^2+2\eta}\Bigr)^2q_i
\label{5.9}\\
\dot\lambda&=&0\,,
\label{5.10}\\
\dot p_\lambda&=&0\,,
\label{5.11}\\
\dot\eta&=&-\,(2\tilde T_1+1)\,
\frac{q^2}{(q^2+2\eta)^2}
+\lambda^2\,q^2\,(p.q-\pi q^2)\approx0\,,
\label{5.12}\\
\dot\pi&=&\frac{2\tilde T_1+1}{q^2+2\eta}
\biggl[\Bigl(\frac{p_i-\pi q_i}{q^2+2\eta}\Bigr)^2q^2-\lambda\biggr]\,
+2(\lambda+\lambda_1)\approx0.
\label{5.13}
\end{eqnarray}

\bigskip\noindent

It is a matter of algebraic work to show that the above equations of
motion  are consistent with the gauge transformations defined in
(\ref{5.4}) in the sense that $[{d\over{dt}},\delta]y=0$  for any
$y$. So gauge transformed variables satisfy the same equations of
motion as the original ones.

\medskip
The combination of the Eqs. (\ref{5.8}) and (\ref{5.9}) leads to


\begin{equation}
\frac{d}{dt}\Bigl(\dot q_i+\lambda^2 q_i\Bigr)
=-\biggl[\lambda^2
+\frac{2\tilde T_2(2\tilde T_1+1)}{(q^2+2\eta)^2}\biggr]
(\dot q_i+\lambda^2 q_i)
-\frac{1}{q^2}(\dot q_k+\lambda^2q_k)^2\,q_i\,.
\label{5.14}
\end{equation}

\bigskip\noindent

\begin{multicols}{2}
As can be verified, the above equation is also consistent with the
gauge transformations. Now, under the constraint surface, it reduces
trivially to

\begin{equation}
\frac{d}{dt}(\dot q_i+\lambda^2 q_i)
=-\lambda^2(\dot q_i+\lambda^2q_i)
-\frac{1}{q^2}(\dot q_k+\lambda^2q_k)^2q_i\,,
\label{5.15}
\end{equation}

\noindent which obviously reproduces (\ref{3.3}) with the gauge
choice $\lambda^2=0$, showing in this way that the partial embedding
procedure introduces gauge degrees of freedom but keeps the same
physical content, as it should be.

\bigskip
To conclude, let us recall the physical meaning of some embedding
results found in literature by using the BFFT formalism. For the
simplest case where the first class constraints of the embedding
theory are linear in the new variables of the extended space, like
chiral-bosons \cite{Floreanini,Miao,Barc1} and Abelian massive vector
theory \cite{BFFT1,BFFT2}, the physical meaning is that the result is
invariant for a shift of coordinates \cite{Amorim}. For the case where
constraints are not linear in these variables, like the non-Abelian
massive vector theory \cite{Barc2}, the embedding theory is
equivalent to the generalized Stuckelberg formalism \cite{Kuni}. The
natural question now is related to the physical meaning of
the result expressed by the equation of motion given by
(\ref{5.15}). We can show that the result above and the initial one
corresponding to a particle on a $N$-sphere described by coordinates
$q_i$ satisfying (\ref{3.3}) are linked by a scale transformation. In
fact, performing a scale transformation over the coordinates $q_i$ in
such a way that

\begin{equation}
Q_i=e^\Lambda q_i\,,
\label{5.16}
\end{equation}

\bigskip\noindent

one directly verifies that (\ref{3.3}) leads to

\begin{equation}
\frac{d}{dt}(\dot Q_i-\dot\Lambda Q_i)=\dot\Lambda(\dot
Q_i-\dot\Lambda Q_i)-\frac{1}{Q^2}(\dot Q_k-\dot\Lambda Q_k)^2Q_i\,.
\label{5.17}
\end{equation}

\bigskip\noindent
By comparing (\ref{5.17}) and (\ref{5.15}) we see that $\lambda^2$
plays the rule of minus the time derivative of the scale factor
$\Lambda$, when the $q$'s of (\ref{5.15}) are interpreted as the
$Q$'s of (\ref{5.17}). This could also be directly seen from Eqs.
(\ref{5.2}) and (\ref{5.5}), namely, $\delta q_i=\epsilon_2q_i$ and
$\delta\lambda^2=-\dot\epsilon_2$. So we see that in some sense the
effect of the embedding procedure on the Hamiltonian system describing
the $N$ dimensional rotor is related to a scale transformation. Fixing
the gauge corresponds to make the scale factor constant over the time,
as expected.

\vspace{1cm}
\section{Conclusion}
In this work we have considered the quantum mechanical analog of the
nonlinear sigma-model, corresponding to a particle constrained to
move on a $N$-dimensional sphere of unit radius. The Hamiltonian
treatment of this model generates four second class constraints. The
embedding of this theory by using the BFFT formalism runs into
difficulties. This is so because it is not natural the choice of
solution in each step of the method and, consequently, one cannot
infer the general rule for higher contributions. This makes the
formalism uninteresting and not feasible to be applied, because we
would have to get an infinite number of corrections to analyze the
embedded theory. On the other hand, we have shown that the same BFFT
method can be conveniently used in order to embed the theory in a
partial way, where just two of the four constraints are converted into
first-class. We have shown that, contrarily to the attempt of using
the full method, all the steps of the corrections are naturally
obtained and, more than that, can be cast in a closed form. Finally,
we have discussed the dynamics and the physical meaning of the
embedded theory. We have shown that it corresponds to a time dependent
scale transformation of coordinates, suggesting us some equivalence
with a geometrical conformal formulation.

\vspace{1cm}
\noindent
{\bf Acknowledgment:} This work is supported in part by Conselho
Nacional de Desenvolvimento Cient\'{\i}fico e Tecnol\'ogico - CNPq,
and Funda\c{c}\~ao Universit\'aria Jos\'e Bonif\'acio - FUJB
(Brazilian Research Agencies).

\end{multicols}

\bigskip
\vspace{1cm}
\begin{thebibliography}{100}
\bibitem[a]{}e-mail: {\tt amorim@ if.ufrj.br}
\bibitem[b]{}e-mail: {\tt barcelos@ if.ufrj.br}
\bibitem[c]{}e-mail: {\tt wotzasek@if.ufrj.br}
\bibitem[]{}
\bibitem{Banerjee} N. Banerjee, R. Banerjee, and S. Ghosh, Nucl. Phys.
B417 (1994) 257; Phys. Rev. D49 (1994) 1996.
\bibitem{BFFT1}I.A. Batalin and E.S. Fradkin, Phys. Lett. B180 (1986)
157; Nucl. Phys.  B279 (1987) 514; I.A. Batalin, E.S. Fradkin, and
T.E. Fradkina, {\it ibid.} B314 (1989) 158; B323 (1989) 734.
\bibitem{BFFT2} I.A. Batalin and I.V. Tyutin, Int. J. Mod. Phys. A6
(1991) 3255.
\bibitem{Dirac} P.A.M.  Dirac, Can. J. Math. 2 (1950) 129; {\it
Lectures on quantum mechanics} (Yeshiva University, New York, 1964).
\bibitem{Floreanini} R. Floreanini and R. Jackiw, Phys. Rev. Lett. 59
(1987) 1873.
\bibitem{Miao} Y.-G. Miao, J.-G. Zhou and Y.-Y. Liu, Phys. Lett. B323
(1994) 169.
\bibitem{Barc1} R. Amorim and J. Barcelos-Neto, Phys. Lett. B333
(1994)
413; Phys. Rev. D53 (1996) 7129.
\bibitem{Amorim} R. Amorim and A. Das, Mod. Phys. Lett. A9 (1994)
3543; R. Amorim, Z. Phys. C67 (1995) 695.
\bibitem{Barc2} R. Banerjee and J. Barcelos-Neto, Nucl. Phys. B499
(1997) 453; M.-I. Park and Y.-J. Park, Int. J. Mod. Phys. A13 (1998)
2179.
\bibitem{Barc3} J. Barcelos-Neto, Phys. Rev. D55 (1997) 2265.
\bibitem{Kuni} T. Kunimasa and T. Goto, Prog. Theor. Phys. 37 (1967)
452.
\end{thebibliography}
\end{document}